\documentclass[preprint,superscriptaddress]{revtex4}
\usepackage{graphicx}
\usepackage[usenames]{color}
\usepackage{amsmath, amsthm, amsfonts,amssymb}

\begin{document}

\title{Effect of changing data size on eigenvalues in the Korean and Japanese stock markets}

\author{Cheoljun Eom}
\affiliation{Division of Business Administration, Pusan National University, Busan 609-735, Republic of Korea}
\email{shunter@pusan.ac.kr}

\author{Woo-Sung Jung}
\affiliation{Department of Physics and Basic Science Research Institute, Pohang University of Science and Technology, Pohang 790-784, Republic of Korea}
\affiliation{Center for Polymer Studies and Department of Physics, Boston University, Boston, MA 02215, USA}

\author{Taisei Kaizoji}
\affiliation{Division of Social Sciences, International Christian University, Tokyo 181-8585, Japan}

\author{Seunghwan Kim}
\affiliation{Department of Physics and Basic Science Research Institute, Pohang University of Science and Technology, Pohang 790-784, Republic of Korea}
\affiliation{Asia Pacific Center for Theoretical Physics, Pohang 790-784, Republic of Korea}

\date{\today}

\begin{abstract}
In this study, we attempted to determine how eigenvalues change, according to random matrix theory (RMT), in stock market data as the number of stocks comprising the correlation matrix changes. Specifically, we tested for changes in the eigenvalue properties as a function of the number and type of stocks in the correlation matrix. We determined that the value of the eigenvalue increases in proportion with the number of stocks. Furthermore, we noted that the largest eigenvalue maintains its identical properties, regardless of the number and type, whereas other eigenvalues evidence different features.
\end{abstract}

\maketitle

\section{Introduction}
Random matrix theory (RMT), which is capable of eliminating random properties from financial time series, has been previously introduced and applied in the field of finance \cite{1,2,3,4}. The RMT employs eigenvalues and eigenvectors to generate a correlation matrix and time series data with various properties. It has been verified that the eigenvalues, which belong to the range beyond the range of the random matrix, bear certain economic implications, such as market factor and industrial factors \cite{2,5,6}. Meanwhile, many studies that have employed the RMT in econophysics are quite similar to studies addressing the deterministic factors of the stock pricing mechanism in the financial field. These studies are also reminiscent of principal component analysis, a multivariate statistical analysis used to examine deterministic factors \cite{7,8,9,10,11}. In the field of finance, these studies have been conducted in combination to develop pricing mechanism models, including the one-, three-, and multi-factor models \cite{12,13,14}. The deterministic factors utilized in each model are the market, industrial, macro-economic, and company factors; these did not differ from the results confirmed by the RMT \cite{2,5,6}.

Identifying the factors that affect the value of the eigenvalue has been an interesting research topic, because the eigenvalue is a crucial parameter not only in finance studies based on multivariate statistics, but also in econophysics studies based on the RMT. As was the case in previous studies, the values of eigenvalues elicited from the financial time series data of various countries differ, and clear differences were determined to exist in the largest eigenvalue. Among the influential factors mentioned thus far in studies involving the RMT, the length of the time series data and the number of stocks influenced the eigenvalue probability density function of the correlation matrix \cite{15}. The findings of finance studies suggested that the largest eigenvalue contributes a large fraction to the variance of returns, and its relative importance increases with the number of stocks more dramatically than others \cite{8,9}. That is to say, the value of the eigenvalue is clearly affected by the number of stocks. The these studies employed multivariate statistics techniques (approximate factor model \cite{10,11}).

In this study, we investigate empirically the relationship between eigenvalues via the RMT and the number of stocks comprising the correlation matrix, as the number of stocks increases. Also, unlike previous studies, we reinforce these results by assessing whether the properties of the eigenvalue change as a function of the numbers and types of stocks within the correlation matrix. We determined that the eigenvalue elicited via the RMT method is directly affected by the variation in the number of stocks in the correlation matrix. On the other hand, the largest eigenvalue maintains its properties regardless of the changes in the numbers and types of stocks in the correlation matrix, whereas other eigenvalues that exceed the range of the random matrix evidence different properties when there were changes in the number and types of stocks. These results suggest that although the largest eigenvalue is affected directly by the number of stocks in the correlation matrix, the properties of the largest eigenvalue do not change.

This paper is constructed as follows. After the introduction, Chapter II provides the data and methods employed in this study. In Chapter III, we show the results obtained in relation to our established research aims. Finally, we summarize the findings and conclusions of this study.

\section{Data and methods}
\subsection{Data}
We evaluated the daily data of stock prices on the Korean and Japanese markets (from Datastream). The stocks were selected via the following process. First, we selected stocks with consecutive daily stock prices for the 18 years from January 1990 to December 2007. Second, the stocks in industry sectors with four or less stocks were excluded. Third, the stocks with extreme outliers, in terms of the descriptive statistics of stock returns, skewness$>|2|$, and kurtosis$>30$, were also excluded. The data selected were $N=358$ stocks from the Korean KOSPI and $N=1099$ stocks from the Japanese TOPIX. The stock returns, $R(t)$, were calculated by the logarithmic changes of the prices $R(t)=\ln P(t)-\ln P(t-1)$, in which $P(t)$ represents the stock price on day $t$.

The number of stocks was determined as follows. The minimum number of stocks in the correlation matrix is set at 50, with an increment of 10. For the Korean market, the number begins at the minimum value of 50 ($=M_1$), and was increased in increments of 10 for 16 rounds, up to 200 ($=M_{16}$). For the Japanese stocks, the number was increased for 36 rounds, up to 400 ($=M_{36}$). In order to minimize the selection bias, 100 iterations were conducted for each number of stocks, and the types of stocks in each iteration are not identical.

\subsection{Random matrix theory}
The RMT was introduced as a method for the control and adjustment of the correlation matrix with measurement errors in a financial time series. According to the statistical properties of the correlation matrix created by the random interactions \cite{15}, if the length of the time series, $L$, and the number of stocks, $N$, is infinite, the eigenvalue, $\lambda$, the probability density function of the correlation matrix, $P_{RM}(\lambda)$, is defined by

\begin{eqnarray}
&&P_{RM}(\lambda)=\frac{Q}{2\pi}\frac{\sqrt{(\lambda_{+}^{RM}-\lambda)(\lambda-\lambda_{-}^{RM})}}
{\lambda} \\
&&\ [ \lambda_\pm^{RM}=1+\frac{1}{Q}\pm 2 \sqrt{\frac{1}{Q}},~~~Q \equiv \frac{L}{N} > 1 ] \nonumber
\end{eqnarray}

\noindent where $\lambda_+^{RM}$ and $\lambda_-^{RM}$ correspond to the maximum and minimum eigenvalues, respectively \cite{15}. We employ eigenvalues in the range beyond the maximum eigenvalue, $\lambda_i>\lambda_+^{RM}$, $i=1,2,\dots,K$, on the basis of the eigenvalue range of the random matrices. In this study, $K=13$ eigenvalues deviated from the random matrix in the Korean stocks, and $K=19$ deviated from the random matrix for the Japanese stocks.

Additionally, in order to determine whether the properties of the eigenvalue change according to changes occurring in the numbers and types of stocks of the correlation matrix, we utilize time series data $R^{E(i)}$ reflective of the properties of each eigenvalue created using the following equation:

\begin{equation}
R_t^{E(i)}=\sum_{j=1}^M V_{i,j}\cdot R_{j,t}
\label{eq:2}
\end{equation}

\noindent where $V_{i,j}$ is the eigenvector of stock $j$ that reflects the $i$th eigenvalue properties, and $R_{j,t}$ is the return of stock $j$ at time $t$. From the correlation matrix of each stock, the time series data of each eigenvalue beyond this range was created from Eq. \ref{eq:2}. Then, via correlation analysis among the created time series data, we attempted to determine whether there was any change in the properties of the eigenvalue, both between and within the number of stocks, respectively.

\section{Result}
\subsection{The Economic Meanings of Eigenvalues}
First of all, we conducted an empirical examination of the economic meanings of eigenvalues that deviated from the range of the random matrix. According to previous studies, these eigenvalue properties have economic meaning, and can function as market, industrial, and macro-economic factors. Because our objective is to determine the effects of eigenvalue properties in accordance with the change in the number of stocks in the correlation matrix, it is necessary to assess whether each eigenvalue does have economic meaning. We created time series data with economic meaning based on the method extensively utilized in finance and econophysics studies, and then examined the relationship between created time series data with economic meanings and those from Eq. (2) in order to reflect the properties of each eigenvalue.

We created the time series data with economic meaning via two methods \cite{6,17}: Equal-weighted returns, $R^{EW}$ and factor scores, $R^{F}$ via factor analysis in multivariate statistics. First, the equal-weighted return is the average return for stocks: $R_t^{EW(q)}=\frac{1}{N_q}\sum_{j=1}^{N_q}R_{j,t}^q$, where $N_q$ represents the number of stocks in the $q$th industry. The overall average return, $N \equiv N_q$, is time series data with market properties, $R_t^{EW(q=1)}$, and the average return for each industry, $N>N_q$, has industrial attributes. There are 14 types of equal-weighted returns, $R_t^{EW(q)}$, $q=1,2,\dots,14$, for the Korean data and 18 types for the Japanese data, including the time series data with market factors, $R_t^{EW(q=1)}$, respectively. Second, in the field of finance, the time series data of deterministic factors of the multi-factor model \cite{14} were created by factor analysis in multivariate statistics \cite{7,8,9,10,11}. Factor analysis, a method that is extensively utilized in the field of social science, can reduce the many variables in the given data set to just a few factors. Via factor analysis, we selected significant factors that are regarded as having economic significance, and created the time series data having the properties of significant factors, which are called factor scores in statistics \cite{17,18}. We rendered factor scores identical to the number of eigenvalues beyond the range of the random matrix. In other words, because 13 eigenvalues in the Korean data deviated from the random matrix, 13 factor scores $R_t^{F(p)}$, $p=1,2,\dots,13$ were ultimately created. For the Japanese data, there were 19 factor scores.

\begin{figure}
\includegraphics[width=1.0\textwidth]{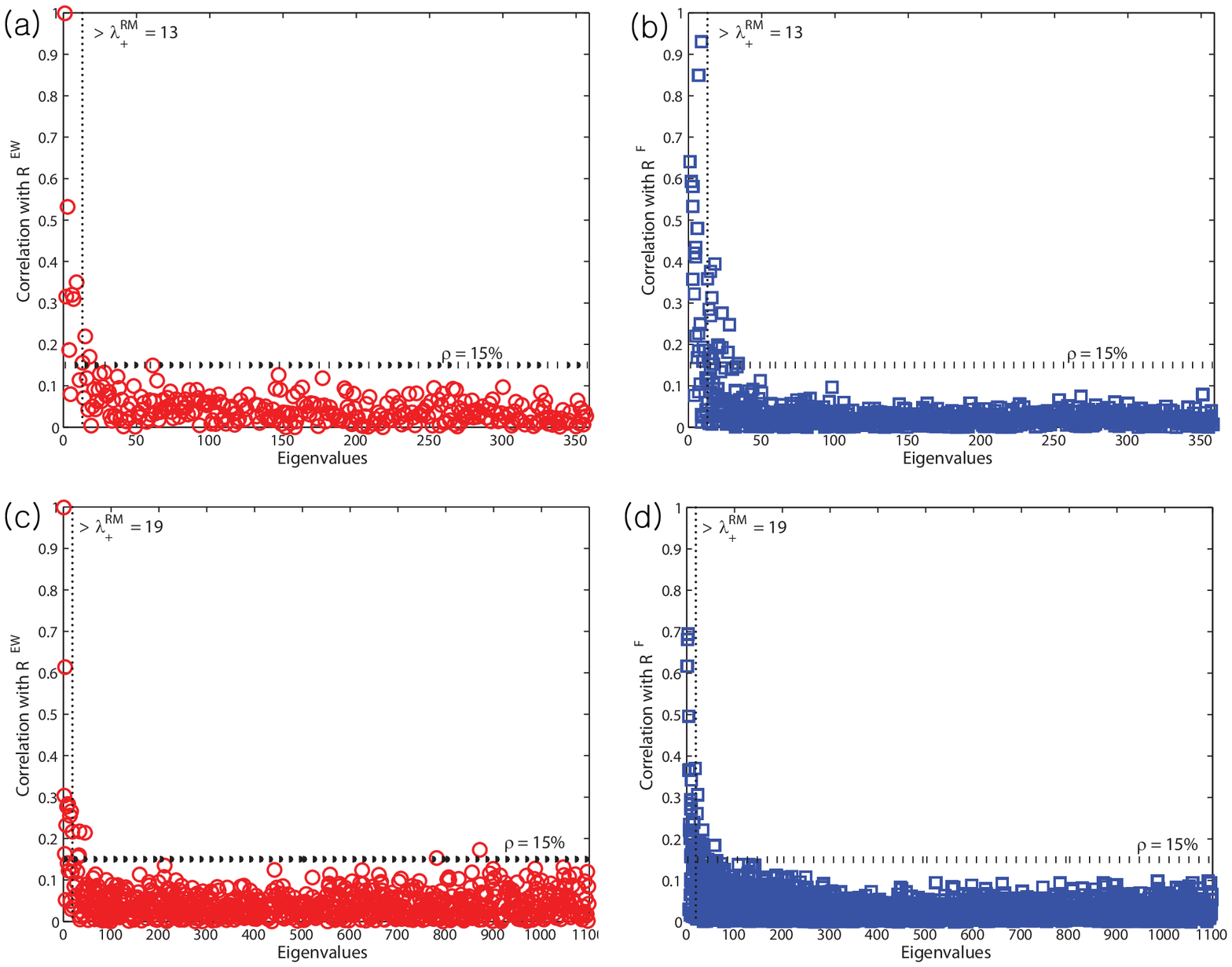}

\caption{(Color online.) The relationship between the time series with economic implications and that from Eq. \ref{eq:2}, which is reflective of the properties of each eigenvalue. In the figure, the X-axis indicates the eigenvalues elicited via the RMT method, and the Y-axis represents the correlation. Fig. 1 (a) \& (c) display the correlation of the maximum values with equal-weighted returns, $R^{EW}$, and Fig. 1(b) $\&$ (d) show the maximum correlation with factor scores, $R^F$. Additionally, Fig. 1 (a) $\&$ (b) depict the results from the Korean data and Fig. 1 (c) $\&$ (d) depict the results from the Japanese data.}
\label{fig:1}
\end{figure}

Fig. \ref{fig:1} presents our findings. The X-axis shows the eigenvalues elicited via the RMT, 358 for Korea and 1,099 for Japan; the Y-axis represents the correlation. Fig. \ref{fig:1}(a) and (c) display the correlation of the maximum values, $\max [\rho (R_t^{E(i)},R_t^{EW(q)})]$, in which $q$ varies from 1 to 14 for Korea and 1 to 18 for Japan, after measuring the correlation between every equal-weighted return and the time series data that reflect the properties of each eigenvalue in Eq. 2. Fig. \ref{fig:1}(b) and (d) show the maximum correlation, $\max [\rho(R_t^{E(i)},R_t^{F(p)})]$, after measuring the correlation between the factor scores created by factor analysis and the time series data from Eq. 2, whereas the $p$ value varies from 1 to 13 for the Korean data and 1 to 19 for the Japanese data. In the figure, the vertical dot-lines denote the maximum eigenvalue, $\lambda_+^{RM}$ in the range of the random matrix, and the horizontal dot-lines represent the benchmark correlation value, $\rho = 15\%$ based on previous studies \cite{7}. Fig. \ref{fig:1}(a) and (b) correspond to the Korean data, and Fig. \ref{fig:1}(c) and (d) are representative of the Japanese data. According to our findings, the eigenvalues beyond the range of the RMT evidence relatively high correlations for equal-weighted return and factor scores, whereas they have very low correlations $\rho < 15\%$ for other eigenvalues. As in previous studies \cite{6,7,8,9,16}, we confirmed empirically that the properties of eigenvalues that deviated from the range of the random matrix had economic implications, including market and industrial factors.

\subsection{The Relationship between Eigenvalues and the number of Stocks}
\begin{figure}
\includegraphics[width=1.0\textwidth]{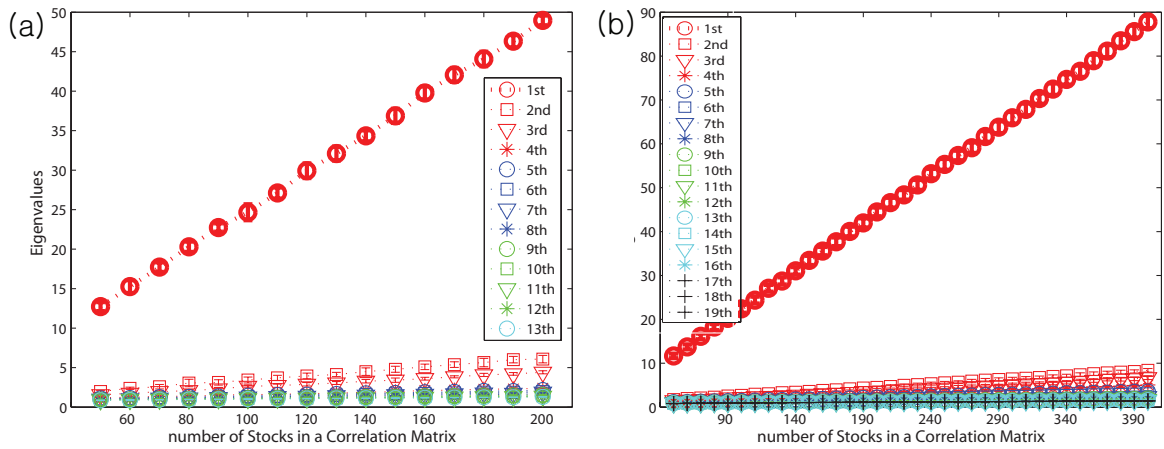}
\caption{(Color online.) The effect on the values of eigenvalues deviated from the random matrix as the number of stocks in the correlation matrix increases. In the figure, the X-axis indicates the number of stocks in the correlation matrix, and the Y-axis represents the eigenvalue. Additionally, Fig. 2(a) shows the results from the Korean data and Fig. 2(b) shows the results from the Japanese data.}
\label{fig:2}
\end{figure}

In this section, we evaluated the effects in the eigenvalues beyond the random range as the number of stocks in the correlation matrix increased. The results are provided in Fig. \ref{fig:2}. The X-axis reflects the number of stocks within the correlation matrix: $M_1, M_2, \dots,M_{16}$ for Korea and $M_1, M_2, \dots,M_{36}$ for Japan. The Y-axis represents the eigenvalue. In order to avoid selection bias, 100 iterations were conducted for each number of stocks, and the types of stocks selected in each iteration were not identical. In the figure, the results are shown in the error-bar in order to represent effectively the observed results of 100 iterations. Fig. \ref{fig:2}(a) corresponds to the Korean data, and Fig. \ref{fig:2}(b) represents the Japanese data. We determined that as the number of stocks in the correlation matrix increases, the value of the eigenvalue increases proportionally. Moreover, we observed from this figure that the largest eigenvalue is significantly greater than the other eigenvalues that deviated from the random matrix. Using these results, we confirmed that the eigenvalues beyond the random range of the RMT were a function of the number of stocks.

Unlike the case in previous studies, we reinforced the observed results by assessing whether the properties of the eigenvalues can be influenced by changes in the numbers and types of stocks in the correlation matrix. In order to investigate this objective, we categorize the relationship between the eigenvalue time series using different numbers of stocks, $\rho^B \equiv \rho \left[ R_{t,M(a)}^{E(i)},R_{t,M(b)}^{E(i)}\right]$, $a\ne b$, and an identical number of stocks, $\rho^W \equiv \rho \left[ R_{t,M(a)}^{E(i)},R_{t,M(b)}^{E(i)}\right]$, $a=b$, respectively. In cases in which there is no change in the eigenvalue properties, the degree of correlation will converge to $\rho \rightarrow 1$. Otherwise, the degree of correlation will approach zero.

\begin{figure}
\includegraphics[width=1.0\textwidth]{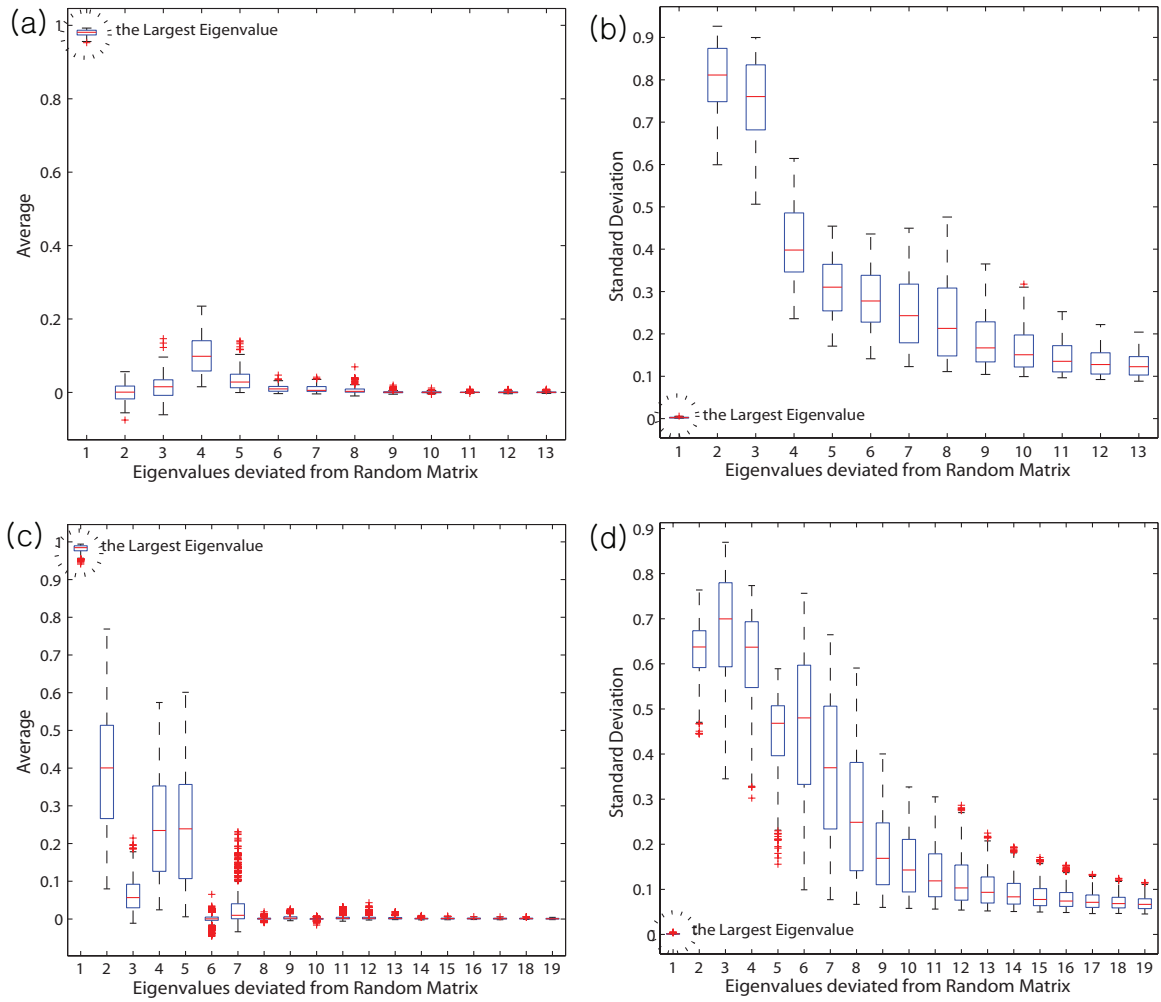}
\caption{(Color online.) The relationship between eigenvalue time series data from \emph{different} numbers of stocks. Fig. 3(a) \& (c) are box-plots of the mean of the correlation, and Fig. 3(b) \& (d) from the standard deviation. In addition, Fig. 3(a) \& (b) depict the results from the Korean data and Fig. 3(c) \& (d) from the Japanese data.}
\label{fig:3}
\end{figure}

First of all, the findings of the relationship between the eigenvalue time series data from \emph{different} number of stocks $\rho^B$ are shown in Fig. \ref{fig:3}. With the total number of stocks $N$ and the number of specific stocks $M$ within a correlation matrix, we selected 100 cases from the possible stock combinations $N!/M!(N-M)!$, without identical types of stocks that comprise the correlation matrix. Accordingly, $k=10,000(=100\times 100)$ correlations were calculated, and we measure the mean $\overline{\rho^{\rm B}} = \frac{1}{10000} \sum_{k=1}^{10000} \rho_k^B$ and the standard deviation $\sigma^{\rm B} = \sqrt{ \sum_{k=1}^{10000} [ \rho_k^B - \overline{\rho^B}]^2 / (10000-1)}$. The number of cases for the calculation of $\overline{\rho^B}$ and $\sigma^B$ were 120 (=$M(M-1)/2 = 16/(16-1)/2$) for Korea and 630 (=$36(36-1)/2$) for Japan, because the measurements were calculated for every number of stocks, from the minimum $M_1$ to the maximum $M_{16}$ of Korea, and $M_{36}$ of Japan. Because 13(19) eigenvalues were beyond the random matrices from the Korean (Japanese) data, the aforementioned testing process was repeated for each of the eigenvalues.

In Fig. \ref{fig:3}, the measured mean and standard deviation are indicated in box-plots. Fig. \ref{fig:3}(a) and (c) correspond to the means of the correlation, and Fig. \ref{fig:3}(b) and (d) represent the standard deviations, and in Fig. \ref{fig:3}(a) and (b) show the results from Korean and Fig. \ref{fig:3}(c) and (d) from Japanese. It was interesting to note that the properties of the largest eigenvalue do not change with the number and types of stocks in a correlation matrix. The mean with the properties of the largest eigenvalue was quite high, $\overline{\rho^B}\geq95\%$ [Fig. \ref{fig:3}(a) \& (c)], but the standard deviation was quite small, $\rho^B\approx 0$ [Fig. \ref{fig:3}(b) \& (d)]. On the other hand, other eigenvalues that deviated from the random matrix have very small mean values with high standard deviation values. This indicates that the change in the eigenvalue properties is extremely sensitive to changes in the numbers and types of stocks.

\begin{figure}
\includegraphics[width=1.0\textwidth]{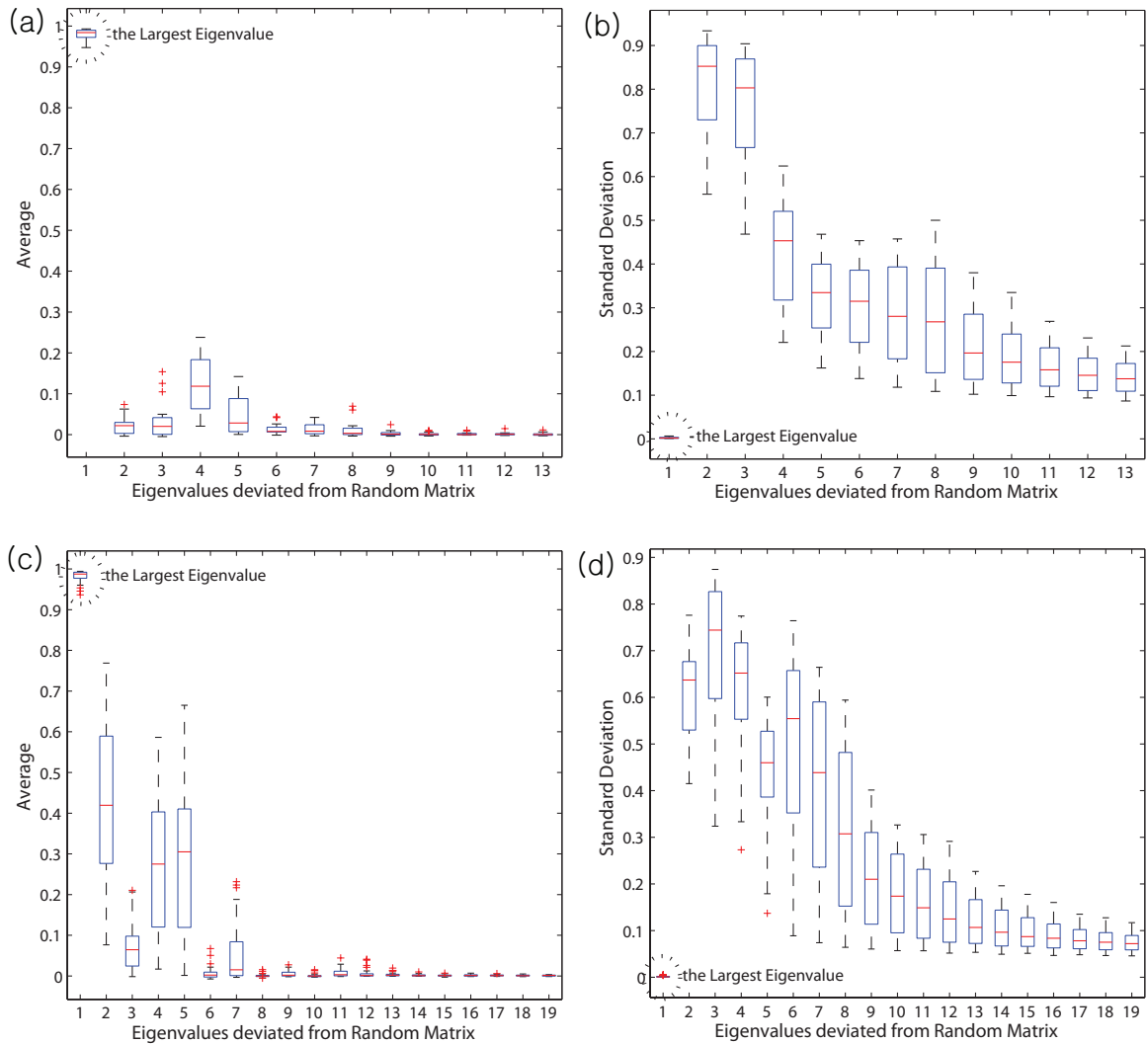}
\caption{(Color online.) The relationship between the eigenvalue time series data from \emph{identical} number of stocks. Fig. 4(a) \& (c) are box-plots of the mean of the correlation, and Fig. 4(b) \& (d) from the standard deviation. Additionally, Fig. 4(a) \& (b) depicts the results from the Korean data and Fig. 4(c) \& (d) from the Japanese data.}
\label{fig:4}
\end{figure}

Next, the findings of the relationship between the eigenvalue time series data from \emph{identical} number of stocks $\rho^W$ are shown in Fig. \ref{fig:4}. We also selected $100$ cases from the possible stock combinations. Accordingly, 4,950(=$100(100-1)/2$) correlations were calculated, and we measured the mean, $\overline{\rho^{W}}$ and the standard deviation, $\sigma^{W}$. The number of cases used to calculate the mean and standard deviation were $M_{16}$ for Korea and $M_{36}$ for Japan; additionally, the aforementioned testing process was repeated for each eigenvalue. In Fig. \ref{fig:4}, the measured mean and standard deviation are shown in box-plots. Fig. \ref{fig:4}(a) and (c) are box-plots of the mean, $\overline{\rho^{W}}$, and Fig. 4(b) and (d) for the standard deviation, $\sigma^{W}$. Fig. \ref{fig:4}(a) and (b) correspond to Korea and Fig. \ref{fig:4}(c) and (d) are representative of Japan. According to the observed results, we determined that the properties of the largest eigenvalue did not change with the type of stocks within a correlation matrix with an identical number of stocks. In other words, the mean correlation among the time series data of the largest eigenvalue is quite high, $\overline{\rho^W}\ge95\%$ [Fig. \ref{fig:4}(a) \& (c)], but the standard deviation is quite small, $\sigma^{\rm W}\approx 0$ [Fig. \ref{fig:3}(b) \& (d)], regardless of the types of stocks in a correlation matrix. On the other hand, other eigenvalues beyond the random range evidence small means and high standard deviation values. This indicates that the eigenvalue properties are sensitive to changes in the type of stocks.

To summarize, we determined herein that even if the value of the eigenvalue elicited via the RMT increases in proportion with the number of stocks in the correlation matrix, the largest eigenvalue maintains its identical properties, regardless of the number and types of stocks in the dataset. However, other eigenvalues evidence different features. The reason for this is as follows. The primary common factor in the field of finance is the market factor that is included in every stock, and the largest eigenvalue has the properties of market factors. Because every stock incorporates market factors regardless of the number and types of stocks, the properties of the largest eigenvalue are not influenced by changes in the number and type of stocks. However, others, including industrial factors, are limited to the stocks in particular industries. Because other eigenvalues have industrial factors, they are extremely sensitive to the numbers and types of stocks. Finally, these findings suggest that studies in which the properties of eigenvalues elicited via the RMT are employed should consider that eigenvalue properties can vary in accordance with the data for eigenvalues other than the largest eigenvalue.

\section{Conclusions}
In the fields of finance and econophysics, the extraction of significant information from the correlation matrix is a fascinating research topic. The field of finance has previously employed multivariate statistics, including principal component analysis, and the RMT was introduced in the field of econophysics. We conducted an empirical study as to how the value of the eigenvalue elicited via the RMT is influenced by the number of stocks in the correlation matrix. Additionally, we reinforced the observed result by assessing whether the properties of the eigenvalues change with the number and types of stocks comprising the correlation matrix.

We determined that the value of the eigenvalue increases in proportion to the number of stocks in the correlation matrix. In particular, the largest eigenvalue increases to a greater degree than the other eigenvalues that deviate from the random matrix. Furthermore, we determined that the largest eigenvalue maintains its identical properties, regardless of the numbers and types of stocks in the correlation matrix. This is attributable to the fact that the properties of the largest eigenvalue are concerned with the market factors incorporated in every stock. However, the properties of other eigenvalues beyond the random range have industrial factors limited to specific stock groups. In this case, the numbers and types of stocks can influence the attributes of each eigenvalue elicited via the RMT.

\begin{acknowledgements}
This work was supported by the Korea Science and Engineering Foundation (KOSEF) grant funded by the Korea government (MEST) (No. R01-2008-000-21065-0), and for two years by Pusan National University Research Grant.
\end{acknowledgements}

\newpage
\clearpage

\end{document}